
\input harvmac

\input amssym.def
\input amssym
\baselineskip 14pt
\magnification\magstep1
\parskip 6pt

\font \bigbf=cmbx10 scaled \magstep1

\newdimen\itemindent \itemindent=32pt
\def\textindent#1{\parindent=\itemindent\let\par=\resetpar%
\indent\llap{#1\enspace}\ignorespaces}

\let\oldpar=\par
\def\resetpar{\oldpar\parindent=20pt\let\par=\oldpar}

\font\ninerm=cmr9 \font\ninesy=cmsy9
\font\eightrm=cmr8 \font\sixrm=cmr6
\font\eighti=cmmi8 \font\sixi=cmmi6
\font\eightsy=cmsy8 \font\sixsy=cmsy6
\font\eightbf=cmbx8 \font\sixbf=cmbx6
\font\eightit=cmti8
\def\eightpoint{\def\rm{\fam0\eightrm}
  \textfont0=\eightrm \scriptfont0=\sixrm \scriptscriptfont0=\fiverm
  \textfont1=\eighti  \scriptfont1=\sixi  \scriptscriptfont1=\fivei
  \textfont2=\eightsy \scriptfont2=\sixsy \scriptscriptfont2=\fivesy
  \textfont3=\tenex   \scriptfont3=\tenex \scriptscriptfont3=\tenex
  \textfont\itfam=\eightit  \def\it{\fam\itfam\eightit}%
  \textfont\bffam=\eightbf  \scriptfont\bffam=\sixbf
  \scriptscriptfont\bffam=\fivebf  \def\bf{\fam\bffam\eightbf}%
  \normalbaselineskip=9pt
  \setbox\strutbox=\hbox{\vrule height7pt depth2pt width0pt}%
  \let\big=\eightbig  \normalbaselines\rm}
\catcode`@=11 %
\def\eightbig#1{{\hbox{$\textfont0=\ninerm\textfont2=\ninesy
  \left#1\vbox to6.5pt{}\right.\n@@space$}}}
\def\vfootnote#1{\insert\footins\bgroup\eightpoint
  \interlinepenalty=\interfootnotelinepenalty
  \splittopskip=\ht\strutbox %
  \splitmaxdepth=\dp\strutbox %
  \leftskip=0pt \rightskip=0pt \spaceskip=0pt \xspaceskip=0pt
  \textindent{#1}\footstrut\futurelet\next\fo@t}
\catcode`@=12 %

\def\a{\alpha}
\def\b{\beta}
\def\c{\gamma}
\def\d{\delta}
\def\e{\epsilon}

\def\l{\lambda}
\def\m{\mu}
\def\n{\nu}
\def\o{\theta}

\def\r{\rho}
\def\s{\sigma}

\def\C{\Gamma}
\def\D{\Delta}

\def\pl{\partial}
\def\rta{\rightarrow}

\def\la{\langle}
\def\ra{\rangle}

\lref\SVeta{G.M. Shore and G. Veneziano, {Nucl. Phys.} {B381} (1992) 3.}
\lref\SVgt{G.M. Shore and G. Veneziano, {Phys. Lett.} {B244} (1990) 75;
{Nucl. Phys.} {B381} (1992) 23.}
\lref\NSVgt{S. Narison, G.M. Shore and G. Veneziano, {Nucl. Phys.} {B433} (1995)
209.}
\lref\NSVgtmass{S. Narison, G.M. Shore and G. Veneziano, {Nucl. Phys.}
{B546} (1999) 235.}
\lref\Smontp{G.M. Shore, {Nucl. Phys. {\it Proc. Suppl.}} {B64} (1998) 167.}
\lref\Szuoz{G.M. Shore, {\it in} Proceedings of the 1998 Zuoz Summer School on
`Hidden Symmetries and Higgs Phenomena', ed. D. Graudenz, 1998; 
hep-ph/9812354.}
\lref\Serice{G.M. Shore, {\it in} Proceedings of the International School of
Subnuclear Physics: `From the Planck Length to the Hubble Radius',
Erice, 1998; hep-ph/9812355.}
\lref\DiVV{P. Di Vecchia and G. Veneziano, {Nucl. Phys.} {B171} (1980) 253.}
\lref\RST{C. Rosenzweig, J. Schechter and G. Trahern, {Phys. Rev.} {D21}
(1980) 3388.}
\lref\Witt{E. Witten, {Nucl. Phys.} {B156} (1979) 269.}
\lref\Veta{G. Veneziano, {Nucl. Phys.} {B159} (1979) 213.}
\lref\Vozi{G. Veneziano, {\it in} `From Symmetries to Strings: Forty Years of
Rochester Conferences', ed. A. Das, World Scientific, 1990.}
\lref\Gusken{S. Gusken, hep-lat/9906034.}
\lref\DiG{G. Boyd, B. All\'es, M. D'Elia and A. Di Giacomo, {\it in} 
Proceedings, HEP97 Jerusalem; hep-lat/9711025\semi
A. Di Giacomo, {\it in} Proceedings, Ahrenshoop Symposium, 1997;
hep-lat/9711034.}
\lref\Nar{S. Narison, {Nucl. Phys.} {B509} (1998) 312.}
\lref\Barc{P. Herrerra-Sikl\'ody, J.I. Latorre, P. Pascual and J. Taron,
{Nucl. Phys.} {B497} (1997) 345; {Phys. Lett.} {B419} (1998) 326\semi
P. Herrerra-Sikl\'ody, {Phys. Lett.} {B442} (1998) 359.}
\lref\KL{H. Leutwyler, {Nucl. Phys. {\it Proc. Suppl.}} {B64} (1998) 223\semi
R. Kaiser and H. Leutwyler, hep-ph/9806336.}
\lref\FKS{T. Feldmann and P. Kroll, {Eur. Phys. J.} {C5} (1998) 327\semi
T. Feldmann, P. Kroll and B. Stech, {Phys. Rev.} {D58} (1998) 11406;
{Phys. Lett.} {B449} (1999) 339\semi
T. Feldmann, {Nucl. Phys. {\it Proc. Suppl.}} {B74} (1999) 151.}
\lref\BFT{P. Ball, J.-M. Fr\`ere and M. Tytgat, {Phys. Lett.} {B365} (1996)
367.}
\lref\EF{R. Escribano and J.-M. Fr\`ere, hep-ph/9901405.}
\lref\Benayoun{M. Benayoun, L. Del Buono, S. Eidelman, V.N. Ivanchenko and
H.B. O'Connell, hep-ph/9902326.}
\lref\PDG{Particle Data Group, C. Caso {\it et al.,} {Eur. Phys. J.}
{C3} (1998) 1.}
\lref\CLEO{CLEO collab., B.H. Behrens {\it et al.,} {Phys. Rev. Lett.}
{80} (1998) 3710.}
\lref\BES{BES collab., J.Z. Bai {\it et al.,} {Phys. Rev.} {D58} (1998) 097101.}
\lref\ELSA{SAPHIR collab., R. Pl\"otzke {\it et al.,} {Phys. Lett.} {B444}
(1998) 555.}
\lref\CEBAF{CEBAF experiments E-89-039 (S. Dytman {\it et al.});
E-91-008 (B.G. Ritchie {\it et al.}).}

{\nopagenumbers
\rightline{CERN--TH/99-233}
\rightline{SWAT/99-234}
\rightline{hep-ph/9908217}
\vskip2cm
\centerline{\bigbf Radiative $\eta'$ Decays, the Topological Susceptibility}
\vskip0.3cm
\centerline{\bigbf and the Witten--Veneziano Mass Formula}
\vskip1cm

\centerline {\bf G.M. Shore}
\vskip0.5cm
\centerline{\it TH Division, CERN,}
\centerline{\it CH1211 Geneva 23, Switzerland}
\vskip0.2cm
\centerline{and}
\vskip0.2cm
\centerline {\it Department of Physics, University of Wales Swansea,}
\centerline {\it Singleton Park, Swansea, SA2 8PP, Wales}

\vskip1cm

{
\parindent 1.5cm{

{\narrower\smallskip\parindent 0pt
The formulae describing the radiative decays $\eta'(\eta)\rta\c\c$
in QCD beyond the chiral limit are derived. The modifications of the
conventional PCAC formulae due to the gluonic contribution to the axial
anomaly in the flavour singlet channel are precisely described.
The decay constants are found to satisfy a modified Dashen formula
which generalises the Witten--Veneziano formula for the mass of the $\eta'$.
Combining these results, it is shown how the topological susceptibility
in QCD with massive, dynamical quarks may be extracted from measurements
of $\eta'(\eta)\rta\c\c$.

\narrower}}}

\vskip2cm

\leftline{CERN--TH/99-233}
\leftline{SWAT/99-234} 
\leftline{July 1999}

\vfill\eject}

\pageno=1

\newsec{Introduction and Results}

The radiative decay $\eta' \rta \c\c$ is of special interest in particle theory
since it involves the gluonic as well as electromagnetic contribution to the
axial anomaly. In this paper, we show how this decay may be used to measure
the topological susceptibility $\chi(0)$ in QCD with dynamical quarks
and thus open a window on the topology of the gluon field.

The flavour non-singlet decay $\pi^0 \rta \c\c$ is of course well understood
and has played an important role in the development of the standard model,
providing early evidence for the existence of colour. Since the pion is
a (pseudo) Goldstone boson for the spontaneously broken chiral symmetry
of QCD, the decay is easily calculated from the electromagnetic contribution
to the anomaly in the corresponding axial current.

However, the theory underlying the flavour singlet decay is less developed
and indeed the special new features arising from the gluon contribution
to the divergence of the $U_A(1)$ axial current are usually still ignored in
phenomenological analyses. In a previous paper\refs{\SVeta}, we presented
an analysis of the theory of $\eta' \rta \c\c$ decay in the chiral limit
of QCD, taking into account the gluonic anomaly and the associated anomalous 
scaling implied by the renormalisation group. Here, we extend that analysis
to QCD with massive quarks, incorporating $\eta - \eta'$ mixing. In
particular, we show how a combination of the radiative decay formula
and a generalisation of the Witten--Veneziano mass formula\refs{\Witt, \Veta} 
for the $\eta'$ 
can be used, under reasonable assumptions, to measure the gluon topological
susceptibility $\chi(0)$ in full QCD with massive quarks.

Our main result is summarised in the following two formulae:
\eqn\Aeq{
f^{a\a}~ g_{\eta^\a\c\c} ~+~ 2n_f A~ g_{G\c\c}~ \d_{a0} ~=~ 
a_{\rm em}^a {\a\over\pi}
}
which describes the radiative decays, and
\eqn\Beq{
f^{a\a} (m^2)_{\a\b} f^{T\b b} = -2 d_{abc} {\rm tr}~ T^c
\left(\matrix{m_u\la\bar u u\ra &0 &0 \cr
0 &m_d\la\bar d d\ra &0 \cr 
0 &0 &m_s\la\bar s s\ra } \right) ~+
(2n_f)^2 A~\d_{a0}\d_{b0}
}
which defines the decay constants appearing in \Aeq\ through a modification
of Dashen's formula to include the gluon contribution to the $U_A(1)$
anomaly. 

In these formulae, $\eta^\a$ denotes the neutral pseudoscalars 
$\pi^0,\eta,\eta'$. The (diagonal) mass matrix is $(m^2)_{\a\b}$ 
and $g_{{\eta^\a}\c\c}$ is the appropriate coupling, 
defined as usual from the decay amplitude by
\eqn\Ceq{
\la \c\c |\eta^\a \ra ~=~ -i ~g_{\eta^\a\c\c}~ \e_{\l\r\a\b}p_1^\a p_2^\b
\e^\l(p_1) \e^\r(p_2)
} 
in obvious notation. The constant $a_{\rm em}^a$ is the coefficient of the
electromagnetic contribution to the axial current anomaly:
\eqn\Deq{
\pl^\m J_{\m5}^a ~=~ d_{acb} m^c \bar q \c_5 T^b q  ~+~ 
2n_f \d_{a0}{\a_s\over8\pi} {\rm tr}G^{\m\n} \tilde G_{\m\n} ~+~ 
a_{\rm em}^a {\a\over8\pi} F^{\m\n}\tilde F_{\m\n}
}
where $a=0,3,8$ is the flavour index and the $d$-symbols are defined from the 
anticommutation relations of the generators, $\{T^a,T^b\} = d_{abc}T^c$.
$G_{\m\n}$ and $F_{\m\n}$ are the gluon and photon field strengths
respectively. (More precise definitions and further notation
are given in sect.2.)

The decay constants $f^{a\a}$ in \Aeq\ are defined by the relation \Beq\ .
Notice immediately that the decay constants which enter the
formula \Aeq\ are {\it not} defined by the coupling of the pseudoscalar
mesons $\pi^0,\eta,\eta'$ to the axial current\refs{\SVeta}. 
In the flavour singlet sector,
such a definition would give a RG non-invariant decay constant which would
not coincide with the quantities arising in the correct decay formula \Aeq\ .
In practice, since flavour $SU(2)$ symmetry is almost exact, the relations
for $\pi^0$ decouple and are simply the standard ones with $f^{3\pi}$
identified as $f_\pi$, viz.
\eqn\Ddeq{
f_\pi g_{\pi\c\c} = {N_c\over3} {\a_{\rm em}\over\pi}
}
together with the Dashen formula
\eqn\Eeq{
f_\pi^2 m_\pi^2 = - (m_u \la\bar u u\ra + m_d \la\bar d d\ra)
}
In the octet-singlet sector, however, there is mixing and the decay constants
form a $2\times 2$ matrix:
\eqn\Feq{
f^{a\a} = \left(\matrix{f^{0\eta'} & f^{0\eta} \cr
f^{8\eta'} & f^{8\eta}} \right) 
}
The four components are independent. In particular, for broken $SU(3)$,
there is no reason to express $f^{a\a}$ as a diagonal matrix times an
orthogonal $\eta - \eta'$ mixing matrix, which would give just three
parameters. Several convenient parametrisations may be made, 
e.g.~involving two constants and two mixing angles\refs{\KL, \FKS,
\EF}, but this does not seem to reflect any special dynamics. 

The novelty of our results of course lies in the extra terms arising in
\Aeq\ and \Beq\ due to the gluonic contribution to the $U_A(1)$ anomaly.
The coefficient $A$ is the non-perturbative number which specifies the 
topological susceptibility in full QCD with massive dynamical quarks.
Defining the topological susceptibility as
\eqn\Geq{
\chi(0) = \int d^4x~i\la0|T~Q(x)~Q(0)|0\ra
}
where $Q = {\a_s\over8\pi}{\rm tr}G^{\m\n} \tilde G_{\m\n}$ is the 
gluon topological charge, the anomalous chiral Ward identities determine 
its dependence on the quark masses and condensates up to an 
undetermined parameter, viz.\refs{\DiVV}
\eqn\Heq{
\chi(0) = {-A~ m_u m_d m_s \la \bar u u\ra \la \bar d d\ra \la \bar s s\ra \over
m_u m_d m_s \la \bar u u\ra \la \bar d d\ra \la \bar s s\ra - A\bigl(
m_u m_d \la \bar u u\ra \la \bar d d\ra +
m_u m_s \la \bar u u\ra \la \bar s s\ra +
m_d m_s \la \bar d d\ra \la \bar s s\ra \bigr)} 
}
or in more compact notation,
\eqn\Hheq{
\chi(0) = -A \biggl(1 - A \sum_q {1\over m_q \la \bar q q\ra}\biggr)^{-1} 
}
Notice how this satisfies the well-known result that $\chi(0)$ vanishes
if any quark mass is set to zero.

The modified Dashen formula is in fact a generalisation of the 
Witten--Veneziano mass formula\refs{\Witt, \Veta} for the $\eta'$. 
Here, however, we do {\it not} impose the leading order in $1/N_c$ 
approximation that produces the Witten--Veneziano formula. 
Recall that, for $n_f=3$ and non-zero quark masses, this states\refs{\Veta}
\eqn\Ieq{
m_{\eta'}^2 + m_{\eta}^2 - 2 m_K^2 =    
-{6\over f_\pi^2} \chi(0)\big|_{\rm YM}
}
To recover \Ieq\ from our result (see the first of eqs(1.13)) the condensate 
$m_s \la \bar s s\ra$ is replaced by the term proportional to $f_\pi^2 m_K^2$ 
using a standard Dashen equation, and the singlet decay constants are set to 
$\sqrt{2n_f}f_\pi$.
The identification of the large $N_c$ limit of the coefficient $A$ with
the non-zero topological susceptibility of pure Yang-Mills theory may be 
seen in different ways, either from the large $N_c$ counting rules
quoted below or perhaps most simply in the effective Lagrangian
analysis explained in section 3. 

The final element in \Aeq\ is the extra `coupling' $g_{G\c\c}$
in the flavour singlet decay formula, which arises because even in the 
chiral limit the $\eta'$ is not a Goldstone boson because of the gluonic
$U_A(1)$ anomaly. A priori, this is {\it not} a physical coupling,
although (suitably normalised) it could be modelled as the coupling of the 
lightest predominantly glueball state mixing with $\eta'$. However,
this interpretation would probably stretch the basic dynamical 
assumptions\foot{The (standard) dynamical assumptions made in
deriving \Aeq\ are described more completely in the following sections.
Essentially, \Aeq\ is based on the {\it zero-momentum} anomalous chiral
Ward identities. It is then assumed that the decay `constants'
$f^{a\a}(k^2)$ and couplings $g_{\eta^\a\c\c}(k^2)$ are approximately 
constant functions of momentum in the range from zero (where the Ward 
identities are applied) to the relevant physical mass. Notice that this is 
only applied to pole-free quantities which depend only implicitly on the 
quark masses. This is of course simply the standard PCAC or chiral 
Lagrangian assumption, and corresponds to assuming pole-dominance of the 
propagators for the appropriate operators by the pseudo Goldstone
bosons. It becomes exact in the chiral limit.}
underlying \Aeq\ too far, and is not necessary
either in deriving or interpreting the formula. In fact, the $g_{G\c\c}$
term arises simply because in addition to the electromagnetic anomaly
the divergence of the axial current contains both the quark bilinear  
operators $\phi_5^a = \bar q \c_5 T^a q$ and the gluonic anomaly $Q$. 
Diagonalising the propagator matrix for these operators isolates the
$\eta$ and $\eta'$ poles, whose couplings to $\c\c$ give the usual
terms $g_{\eta\c\c}$ and $g_{\eta'\c\c}$. However, the remaining
operator (which we call $G$ -- see following sections) also couples
to $\c\c$ and therefore also contributes to the decay formula, whether
or not we assume that its propagator is dominated by a `glueball' pole.

Of course, the presence of the in general unmeasurable coupling $g_{G\c\c}$
in \Aeq\ appears to remove any predictivity from the $\eta'\rta\c\c$ decay 
formula.
In a strict sense this is true, but we shall argue below that it may
nevertheless be a good dynamical approximation to assume $g_{G\c\c}$ is 
small compared to $g_{\eta'\c\c}$. In this case, we can combine eqs.\Aeq\
and \Beq\ to give a measurement of the non-perturbative coefficient $A$ in
$\chi(0)$. Assume flavour $SU(2)$ is exact so that eqs.\Eeq\ and \Feq\
for the pion decouple. The remaining parts of \Aeq\ and \Beq\ then together
provide five equations (since \Beq\ is symmetric), in which we assume the 
physical quantities $m_{\eta}, m_{\eta'}, g_{\eta\c\c}$ and $g_{\eta'\c\c}$
are all known and we neglect $g_{G\c\c}$.
Four of these equations may be used to determine the four decay constants
$f^{a\a}$. The final equation is the flavour singlet Dashen formula, which
may then be solved for $A$. This is the generalisation of the Witten--Veneziano
formula.

Without neglecting $g_{G\c\c}$, the five equations give a self-consistent
description of the radiative decays, but are non-predictive. It is 
therefore important to analyse more carefully whether it is really
legitimate to neglect $g_{G\c\c}.$\foot{Notice that this approximation
is implicitly made in almost all the phenomenological analyses of
radiative pseudoscalar decays, which use a formula like \Aeq\ but omitting
the $g_{G\c\c}$ term and (mistakenly) implying that the decay constants
are defined as the couplings of the axial current to $\eta$ and $\eta'$.}
The argument is based on the fact that $g_{G\c\c}$ is both OZI suppressed 
{\it and} renormalisation group (RG) invariant\refs{\SVeta}. 
Since violations of the OZI 
rule\foot{The OZI approximation to QCD may be given a precise definition
\refs{\Vozi} as the truncation of full QCD in which non-planar and
quark-loop diagrams are retained, but the diagrams which give purely
gluonic intermediate states (those in which the external currents are
attached to different quark loops) are omitted. This is a closer 
approximation to QCD than, for example, the leading large $N_c$ limit.} are
associated with the $U_A(1)$ anomaly, it is a plausible conjecture
that we can identify OZI-violating quantities by their dependence on the 
anomalous dimension associated with the non-trivial renormalisation of 
$J_{\m5}^0$ due to the anomaly. In this way, RG non-invariance can be used 
as a flag to indicate those quantities expected to show large OZI violations.
If this conjecture is correct, then we would expect the OZI rule to be
reasonably good for the RG invariant $g_{G\c\c}$, which would therefore
be suppressed relative to $g_{\eta'\c\c}.$\foot{On the other hand, we
would {\it not} expect the RG non-invariant and anomaly sensitive `decay
constant' $\hat f^{0\eta'}$ defined by $\la 0|J_{\m5}^0|\eta'\ra =
ik_\m \hat f^{0\eta'}$ to be well-approximated by its OZI value.} 
(An important exception is of course the $\eta'$ mass itself, which
although obviously RG invariant is not zero in the chiral limit 
as it would be in the OZI limit of QCD.) Notice 
that this conjecture has been applied already with some success to the
`proton spin' problem in polarised deep inelastic scattering\refs{\SVgt, 
\NSVgt, \NSVgtmass}.

It is also interesting to look at eqs.\Aeq\ and \Beq\ from the point of view
of the large $N_c$ expansion. The large $N_c$ counting for the various
quantities involved is as follows: $f^{a\a} = O(\sqrt{N_c})$, 
$g_{\eta^\a\c\c} = O(\sqrt{N_c})$, $g_{G\c\c} = O(1)$, 
$m_{\eta'}^2 = O(1/N_c)$, $m_{\eta}^2 = O(1)$, $\la \bar q q\ra = O(N_c)$,
$a_{\rm em}^a = O(N_c)$ and $A = O(1)$. 
Notice first that this implies $\chi(0) \simeq A$ in the large $N_c$ limit, 
as already used above to derive the Witten--Veneziano formula. 
It also follows that the term in the decay formula involving $A g_{G\c\c}$
is suppressed by $O(1/N_c)$ relative to the $f^{a\a} g_{\eta^a\c\c}$ term.
If large $N_c$ is reliable in this case, we would therefore indeed
expect this contribution to be suppressed. However, the large $N_c$ limit
must be used with great caution in the $U_A(1)$ channel. For example,
the same argument would equally imply that the additional term $A$ in the
flavour singlet Dashen formula is suppressed by $O(1/N_c)$, yet we know
that although formally of $O(1/N_c)$, $m_{\eta'}^2$ is not small
phenomenologically.

To make the phenomenological application of our results \Aeq\ and \Beq\ quite
clear, we now write out the five equations in the $\eta - \eta'$ sector
explicitly. Set $n_f=3$ and take $m_u = m_d = 0$ for simplicity.
The decay equations are:
\eqn\Jeq{\eqalign{
{}&f^{0\eta'} g_{\eta'\c\c} + f^{0\eta} g_{\eta\c\c}
+ 6 A g_{G\c\c} = a_{\rm em}^0 {\a\over\pi} \cr
{}&f^{8\eta} g_{\eta\c\c} + f^{8\eta'} g_{\eta'\c\c}
= a_{\rm em}^8 {\a\over\pi} \cr
}}
where $a_{\rm em}^0 = {4\over3} N_c$ and $a_{\rm em}^8 = {1\over3\sqrt3} N_c$,
and the Dashen equations are:
\eqn\Keq{\eqalign{
\bigl(f^{0\eta'}\bigr)^2 m_{\eta'}^2 +
\bigl(f^{0\eta}\bigr)^2 m_{\eta}^2 &= -4m_s \la\bar s s\ra + 36 A \cr
f^{0\eta'} f^{8\eta'} m_{\eta'}^2 +
f^{0\eta} f^{8\eta} m_{\eta}^2 &= {4\over\sqrt3} m_s \la\bar s s\ra \cr
\bigl(f^{8\eta}\bigr)^2 m_{\eta}^2 +
\bigl(f^{8\eta'}\bigr)^2 m_{\eta'}^2 &= -{4\over3}m_s \la\bar s s\ra  \cr
}}
Clearly, the two purely octet formulae can be used to find $f^{8\eta}$
and $f^{8\eta'}$ if both $g_{\eta\c\c}$ and $g_{\eta'\c\c}$ are known.
The off-diagonal Dashen formula then expresses $f^{0\eta}$ in terms of
$f^{0\eta'}$.
This leaves the two purely singlet formulae involving the still-undetermined
decay constant $f^{0\eta'}$, the topological susceptibility coefficient $A$,
and the coupling $g_{G\c\c}$.
The advertised result follows immediately. If we neglect $g_{G\c\c}$,
we can find $f^{0\eta'}$ from the singlet decay formula and thus
determine $A$ from the remaining, generalised Witten--Veneziano, formula.

On the other hand, we may regard $A$ as a number to be predicted by
non-perturbative theoretical calculations. In that case, the  
Dashen formula determines the decay constant $f^{0\eta'}$ in terms of $A$,
in which case everything is known in the singlet decay formula except
the coupling $g_{G\c\c}$, which is therefore predicted. Unfortunately, it
is not at all clear how this could be compared to an experimental
measurement without invoking `glueball dominance' in the $\la G~ G\ra$ 
propagator and assuming that this dominates over other neglected
pseudoscalar poles in the propagator matrix (see section 3).

Finally, we should discuss briefly whether it is possible to determine
any of the quantities $f^{a\a}$ or $A$ by non-perturbative calculations,
either using QCD spectral sum rules or lattice gauge theory.
The decay constant definition \Beq\ has an alternative (prior) form in terms
of the propagators for the pseudoscalar quark bilinears $\phi_5^a$.
In fact (for notation and the  derivation see sections 2 and 3), we have:
\eqn\Leq{
f^{a\a} (m^2)_{\a\b} f^{T\b b} ~=~ d_{ace}\la\phi^e\ra ~\la \phi_5 ~\phi_5
\ra_{cd}^{-1}~d_{dbf}\la\phi^f\ra
}
where on the r.h.s.~$\la \phi_5 ~\phi_5 \ra_{cd}^{-1}$ denotes the $(cd)$
component of the inverse of the matrix of two-point functions of the
pseudoscalars, taken at zero momentum, and $\la\phi^a\ra$ are the usual
quark condensates. Comparing with \Beq\ shows that
a successful calculation of the r.h.s.~would imply a determination
of the coefficient $A$ governing the topological susceptibility.
Although this looks relatively straightforward, perhaps unsurprisingly
it turns out to be a very delicate calculation indeed in the
QCD spectral sum rule approach,\foot{I would like to thank Stephan Narison
for collaborating on a preliminary investigation of this problem.} 
primarily because the effects
of gluons and the anomaly have to make an important contribution to 
what are in first approximation purely quark bilinear propagators. 
This may nevertheless be an interesting problem to pursue
or to study on the lattice.

The other possibility is to determine $A$ by a direct calculation of 
the topological susceptibility $\chi(0)$ in full QCD with dynamical,
massive quarks\foot{See refs.\refs{\DiG, \Gusken} 
for reviews of progress in calculating the topological susceptibility in 
full QCD on the lattice. Ref.\refs{\Nar} contains results and references 
relevant to the spectral sum rule approach. 
(See also \refs{\NSVgt , \NSVgtmass}.)}. 
Because of the intricate dependence on all the
quark masses (including the light quarks) this does not seem feasible
using spectral sum rules. The situation may be better on the lattice,
however, if the possibility to change quark masses is exploited. 
Perhaps $A$ could be extracted from a calculation of $\chi(0)$
in the limit of equal, but not too light, quark masses.

A general point highlighted once again by this is that it would be extremely
useful to discover a calculational method that yields 1PI vertices
directly rather than deducing them by amputation of the related
Green functions. For example, as we show in section 2, the coefficent
$A$ in the topological susceptibility is in fact just
given by the `two-point vertex' functional $\C_{QQ}$. These `1PI'
functionals are smooth and free of all the singularities and delicate mass
dependence associated with external propogator poles. They are therefore 
the essential non-perturbative quantities we need to find.

It would also be interesting to make a detailed comparison of the formulae
presented here with the corresponding results in the recently constructed
chiral Lagrangians in which the $\eta'$ is incorporated in the framework of
the $1/N_c$ expansion\refs{\KL, \Barc}. Although the
formalisms look rather different, mainly because of the non-linear
realisations used in the chiral Lagrangian approach and the explicit 
reliance on the $1/N_c$ expansion, the essential dynamical input and 
assumptions are the same and the final physical predictions should 
agree within the limitations of the approximations. 

\vskip0.3cm
The rest of this paper is devoted to the derivation of the above results by
different methods. In section 2 we establish some convenient notation and
review the basic anomalous chiral Ward identities which are the basis of 
all the subsequent work. Section 3 gives a derivation of \Aeq\ and \Beq\ using 
the method of 1PI vertex functionals used to analyse $\eta' \rta \c\c$ in the
chiral limit in refs{\SVeta}. Then in section 4 we write an effective
Lagrangian (extending refs.\refs{\DiVV, \RST}) 
which incorporates all the constraints of 
the zero-momentum chiral Ward identities and includes the coupling to
electromagnetism. In section 5 we give a slightly simplified
derivation which follows as closely as possible the traditional PCAC
methods, generalised as necessary to take account of the gluonic $U_A(1)$
anomaly. This section is intended to be as self-contained as possible,
and readers interested simply in the phenomenology may prefer to go
directly to section 5.

\vskip0.2cm
Finally, the methods of this paper may equally be applied to other decays 
involving the $\eta'$ \refs{\BFT , \EF , \Benayoun}. Of special interest are, 
for example, $\eta'\rta V\c$ where $V$ is a light $1^-$ meson such as $\rho$, 
which is related to $\eta'\rta\c\c$ by vector meson dominance;
$\eta' \rta \pi\pi\c$, which is determined by the box anomaly for 
one vector and three axial currents; and $\psi \rta \eta' \c$ \refs{\PDG}.
Some interesting current experimental studies of $\eta'$ physics include, 
for example, $B\rta \eta' K$ by the CLEO collaboration\refs{\CLEO}; 
$\psi_{2S} \rta \eta'\c$ by the BES collaboration at BEPC\refs{\BES}; 
and $\eta'$ photoproduction at ELSA\refs{\ELSA} and CEBAF\refs{\CEBAF}.

\newsec{Chiral Ward Identities}

The anomalous chiral Ward identities for QCD with massive quarks have
been written down in the form used here in ref.\refs{\NSVgtmass} and reviewed
in \refs{\Szuoz}. We refer to these papers for more complete derivations
and in this section simply define our notation and quote the essential
identities. In this section, we omit the electromagnetic contributions.

The composite operators involved in the Green functions and 1PI vertices
studied here are the currents and pseudoscalar operators 
$J_{\m5}^a$, $Q$, $\phi_5^a$ and the scalar $\phi^a$ where
\eqn\AAeq{\eqalign{
J_{\m5B}^a &= \bar q \c_\m \c_5 T^a q~~~~~~~~
Q_B = {\a_s\over8\pi} {\rm tr} G_{\m\n} \tilde G^{\m\n}~~~~~~~~~
\cr 
\phi_{5B}^a &= \bar q \c_5 T^a q ~~~~~~~~~~
\phi_B^a = \bar q T^a q \cr
}}
$G_{\m\n}$ is the gluon field strength.
In this notation, $T^i = {1\over2}\l^i$ are flavour $SU(n_f)$ generators,
and we include the singlet $U_A(1)$ generator $T^0 = {\bf 1}$ and let the
index $a = 0, i$.
We only need to consider fields where $i$ corresponds to a generator
in the Cartan sub-algebra, so that $a = 0, 3, 8$ for $n_f = 3$ quark
flavours. $d$-symbols are defined by $\{T^a,T^b\} = d_{abc} T^c$. Since this
includes the flavour singlet $U_A(1)$ generator, they are only symmetric on
the first two indices. For $n_f = 3$, the explicit values are $d_{000} =
d_{033} = d_{088} = 2, d_{330} = d_{880} = 1/3, d_{338} = d_{383} =
-d_{888} = 1/\sqrt3$. 

We also use the following compact notation.
The quark mass matrix is written as $m^a T^a$, so that for $n_f=3$, 
\eqn\BBeq{
\left(\matrix{m_u &0 &0 \cr
0 &m_d &0 \cr
0 &0 &m_s \cr}\right)
= m^0 {\bf 1} + m^3 T^3 + m^8 T^8
}
In the same way, the chiral symmetry breaking condensates may be written as 
\eqn\CCeq{
\left(\matrix{ \langle \bar u u\rangle &0 &0 \cr 0 &\langle \bar d d\rangle
&0 \cr
0 &0 &\langle \bar s s\rangle \cr}\right) = {1\over3} \la\phi^0\ra {\bf 1} + 2
\la\phi^3\ra T^3 + 2 \la\phi^8\ra T^8 
}
where $\langle \phi^c\rangle$ is the VEV $\langle \bar q T^c q\rangle$.
We then define 
\eqn\DDeq{
M_{ab} = d_{acb} m^c
~~~~~~~~~~~~~~~~~~
\Phi_{ab} = d_{abc} \langle \phi^c\rangle 
}

Eq \AAeq\ defines the bare operators. The renormalised composite
operators are defined as follows:
\eqn\EEeq{\eqalign{
&J_{\m5}^0 = Z J_{\m5B}^0  ~~~~~~~~~~
J_{\m5}^{a\neq0} =  J_{\m5B}^{a\neq0} \cr
&Q = Q_B - {1\over 2n_f}(1-Z) \pl^\m J_{\m5B}^0 \cr
&\phi_{5}^a = Z_\phi \phi_{5B}^a ~~~~~~~~~~~~~
\phi^a = Z_\phi \phi_B^a \cr
}}
where $Z_\phi$ is the inverse of the mass renormalisation, 
$Z_\phi = Z_m^{-1}$. The non-trivial renormalisation of $J_{\m5}^0$
means that its matrix elements scale with an anomalous dimension $\c$
related to $Z$. This occurs because $J_{\m5}^0$ is not a conserved
current, due to the anomaly $Q$.
Notice also the mixing of the operator $Q$ with $\pl^\m J_{\m5}^0$ under
renormalisation.

The Green functions for these operators are constructed by functional
differentiation from the generating functional\refs{\Szuoz} 
$W[V_{\m5}^a, \o, S_5^a, S^a]$, where
$V_{\m5}^a, \o, S_5^a, S^a$ are the sources for the composite
operators $J_{\m5}^a, Q, \phi_5^a, \phi^a$ respectively.
For example, the Green function $i\la0|T~Q(x)~Q(y)|0\ra$ is given by
${\d^2 W\over \d\o(x) \d\o(y)}$, which we abbreviate as $W_{\o\o}$.
In this notation, the anomalous zero-momentum chiral Ward identities 
are:
\eqn\FFeq{\eqalign{
&2n_f \d_{a0} W_{\o \o} + M_{ac} W_{S_5^c \o} = 0 \cr
&2n_f \d_{a0} W_{\o S_5^b} + M_{ac} W_{S_5^c S_5^b} + \Phi_{ab} = 0 \cr
}}
which implies the following identity for the topological susceptibility,
\eqn\GGeq{
(2n_f)^2 \chi(0) = M_{0c} W_{S_5^c S_5^d} M_{d0} + (M\Phi)_{00}
}
These are derived from the fundamental anomalous Ward identity
\eqn\HHeq{
\pl_\m W_{V_{\m5}^a} - 2n_f \d_{a0} W_{\o} - M_{ac} W_{S_5^c} + d_{adc}
S^d W_{S_5^c} - d_{adc} S_5^d W_{S^c}
= 0
}
which is the precise expression in the functional formalism of the
identity \Deq\ .

The 1PI vertices used in section 3 are defined as functional derivatives
of a second generating functional (effective action) $\C$, constructed
from $W$ by a partial Legendre transform with respect to the fields
$Q$, $\phi_5^a$ and $\phi^a$ only ({\it not} the currents
$J_{\m5}^a$)\refs{\Szuoz}.
The resulting vertices are `1PI' w.r.t.~the propagators for these composite 
operators only. This separates off the particle poles in these 
propagators, and gives the closest identification of the field theoretic
vertices with the physical couplings such as $g_{\eta^\a \c\c}$.

The basic anomalous chiral Ward identity for $\C$ follows immediately
from \HHeq\ for $W$:
\eqn\IIeq{
\pl_\m \C_{V_{\m5}^a} - 2n_f \d_{a0} Q - M_{ac} \phi_5^c + d_{acd}
\phi^d \C_{\phi_5^c} - d_{acd} \phi_5^d \C_{\phi^c} = 0 
}
and other identities follow simply by functional differentiation.
In particular, for the two-point vertices, we find the following
zero-momentum identities (analogous to \FFeq\ ):
\eqn\IIieq{\eqalign{
\Phi_{ac}\C_{\phi_5^c Q} - 2n_f \d_{a0} &= 0 \cr
\Phi_{ac} \C_{\phi_5^c \phi_5^b} - M_{ab} &= 0 \cr
}}
which together imply
\eqn\IIiieq{
\Phi_{ac} \C_{\phi_5^c \phi_5^d} \Phi_{db} = - (M\Phi)_{ab}
}
These will be useful in section 4.

The fact that the topological susceptibility is zero for vanishing quark mass
can be seen immediately from \GGeq\ . One of the simplest ways to derive
the precise form \Heq\ or \Hheq\ is in fact to use an identity involving
$\C$. As is well-known, the two-point vertices are simply the inverse
of the two-point Green functions (propagators), so in the pseudoscalar sector
we have the following matrix inversion formula:
\eqn\JJeq{\eqalign{
\C_{QQ} &= - \Bigl(W_{\o\o} - W_{\o S_5^a} (W_{S_5 S_5})_{ab}^{-1} 
W_{S_5^b \o} \Bigr)^{-1} \cr
&= - \Bigl(W_{\o\o} - W_{\o S_5^a} M_{ac} 
\bigl( M W_{S_5 S_5} M \bigr)_{cd}^{-1} 
M_{db} W_{S_5^b \o} \Bigr)^{-1} \cr
}}
and using the identities \FFeq\ and \GGeq\ this implies
\eqn\KK{
\C_{QQ}^{-1} = - \chi \Bigl(1 - (2n_f)^2 \chi (M\Phi)_{00}^{-1} \Bigr)^{-1}
}
all at zero momentum.
Inverting this relation gives the result for the topological susceptibility:
\eqn\LLeq{
\chi = - \C_{QQ}^{-1} \Bigl(1 - (2n_f)^2 \C_{QQ}^{-1}
(M\Phi)_{00}^{-1} \Bigr)^{-1}
}
Substituting the explicit expression for $(M\Phi)_{00}^{-1}$ (which is easily
found from the definitions above), viz.
\eqn\MMeq{
(M\Phi)_{00}^{-1} = {1\over (2n_f)^2} \sum_q {1\over m_q \la \bar q q\ra}
}
we see that \LLeq\ reproduces the general form \Hheq\ where we can
now identify the (mass-independent) non-perturbative coefficient as
\eqn\NNeq{
A =  \C_{QQ}^{-1}
}

\newsec{$\eta' \rta \c\c$ from 1PI Vertices}

In this section, we present the most theoretically complete derivation
of the decay formula \Aeq\ and generalised Dashen formula \Beq\ . This
follows the derivation previously given in ref.\refs{\SVeta} for the 
chiral limit, extended to include quark masses. The technique relies on the
identification of the couplings $g_{\eta^\a \c\c}$ with the zero-momentum 
limit of certain 1PI vertex functions, precisely defined using the Legendre
transform $\C$ introduced in section 2. These techniques have also been used 
in our series of papers on the $U_A(1)$ Goldberger--Treiman relation
and the `proton spin' \refs{\SVgt, \NSVgt, \NSVgtmass}. See also
\refs{\Smontp, \Serice} for reviews.

The starting point is the Ward identity \IIeq\ extended to include the
electromagnetic contribution to the anomaly for the axial current:
\eqn\AAAeq{
\pl_\m \C_{V_{\m5}^a} - 2n_f \d_{a0} Q 
- a_{\rm em}^a Q_{\rm em}(A) - M_{ac} \phi_5^c + d_{acd}
\phi^d \C_{\phi_5^c} - d_{acd} \phi_5^d \C_{\phi^c} = 0 
}
where $Q_{\rm em}(A)$ is shorthand notation for ${\a\over8\pi} F_{\m\n}
\tilde F^{\m\n}$, where $F_{\m\n}$ is the field strength for the 
electromagnetic field $A_\m$. (Since we are working only to leading order in
$\a$, it is not necessary to consider $Q_{\rm em}$ as an independent
composite operator with non-trivial renormalisation.)

Differentiating twice w.r.t.~the field $A_\m$, evaluating at the VEVs,
and taking the Fourier transform, we find
\eqn\BBBeq{
ik_\m \C_{V_{\m5}^a A^\l A^\r} + a_{\rm em}^a {\a\over\pi} \e_{\l\s\a\b}
p_1^\a p_2^\b + d_{abc} \phi^c \C_{\phi_5^b A^\l A^\r} = 0
}
where $p_1,p_2$ are the momenta of the photons. To simplify notation, 
it will be convenient from now on to define vertices 
$\hat \C$ with the kinematical factors removed, in particular
$\C_{\phi_5^a A^\l A^\r} = -\hat\C_{\phi_5^a A^\l A^\r} 
\e_{\l\s\a\b} p_1^\a p_2^\b$. Notice that the mass
term in \AAAeq\ does not contribute explicitly to this formula.
From its definition as 1PI w.r.t.~the pseudoscalar fields, the vertex
$\C_{V_{\m5}^a A^\l A^\r}$ has no pole at $k^2=0$ (even in the chiral limit)
so the first term vanishes at zero momentum $k$, leaving simply
\eqn\CCCeq{
\Phi_{ab} \hat\C_{\phi_5^b A^\l A^\r}\Big|_{k=0} = a_{\rm em}^a {\a\over\pi}
}

The first step in converting \CCCeq\ to the decay formula \Aeq\
is to identify the physical states $\eta^\a$. These appear as poles in the
propagator matrix for the four pseudoscalar operators $Q$, $\phi_5^a$
($a=0,3,8$). To isolate these poles, we diagonalise the propagator
matrix in this sector then normalise the three operators coupling to the 
physical states.

We therefore define the operator 
\eqn\DDDeq{
G = Q - W_{\o S_5^a} (W_{S_5 S_5})_{ab}^{-1} \phi_5^b
}
so that by construction the propagators $\la G~ \phi_5^a\ra$ all vanish.
(Notice that integrations over repeated spacetime arguments are implied
in this condensed notation.) Then define operators
\eqn\EEEeq{
\eta^\a = C^{\a b} \phi_5^b
}
such that the propagator matrix 
\eqn\FFFeq{
\la \eta^\a~ \eta^\b\ra \equiv W_{S_5^\a S_5^\b}
= C^{\a a} W_{S_5^a S_5^b} C^{T b \b}
= \left(\matrix{{-1\over k^2 - m_{\eta'}^2} &0 &0 \cr
0&{-1\over k^2 - m_{\eta}^2} &0 \cr
0 &0 &{-1\over k^2 - m_{\pi}^2}}\right)
}
where $S_5^\a$ are the sources for the operators $\eta^\a$.

This change of variable affects the partial functional derivatives 
in $\hat\C_{\phi_5^a A^\l A^\r}$ in \CCCeq\ , which involves 
${\d\over\d \phi_5^a}$ at fixed $Q$.
In terms of the new variables $G$, $\eta^\a$ we have
\eqn\GGGeq{\eqalign{
{\d\over\d \phi_5^a}\bigg|_{Q} &= 
{\d\eta^\a\over\d \phi_5^a}{\d\over\d \eta^\a}
+ {\d G\over\d \phi_5^a} {\d\over\d G} \cr
&= C^{T a\a} {\d\over\d \eta^\a}
- (W_{S_5 S_5})_{ab}^{-1} W_{S_5^b \o}{\d\over\d G} \cr
}}
The decay formula therefore becomes
\eqn\HHHeq{
\Phi_{ab} C^{T b\a} ~\hat \C_{\eta^\a A^\l A^\r} -
\Phi_{ab} (W_{S_5 S_5})_{ab}^{-1} W_{S_5^b \o} ~\hat \C_{G A^\l A^\r}
= a_{\rm em}^a {\a\over\pi}
}

The decay constants are identified as
\eqn\IIieq{
f^{a\a} = \Phi_{ab} C^{T b\a}
}
In terms of the propagators, we can write (from \FFFeq\ )
\eqn\IIIeq{
f^{a\a} (W_{S_5 S_5})_{\a\b}^{-1} f^{T \b b} = 
\Phi_{ac} (W_{S_5 S_5})_{cd}^{-1} \Phi_{db}
}
and so at zero momentum
\eqn\JJJeq{
f^{a\a} m_{\a\b}^2 f^{T \b b} = 
\Phi_{ac} (W_{S_5 S_5})_{cd}^{-1} \Phi_{db}
}
as quoted in \Leq\ .

The remaining steps in finding \Aeq\ and \Beq\ are an exercise in manipulating 
the zero-momentum Ward identities \FFeq\ . First note that combining the
two identities in \FFeq\ gives
\eqn\KKKeq{
M_{ac} W_{S_5^c S_5^d} M_{db} = - (M\Phi)_{ab} + (2n_f)^2 \chi(0) 
\d_{a0} \d_{b0}
}
whose $a,b=0$ component is just \GGeq\ . Note that $(M\Phi)_{ab}$ is
symmetric. Also define ${\bf 1}_{00}=\d_{a0} \d_{b0}$. Then we can write
\eqn\LLLeq{\eqalign{
\Phi_{ab} (W_{S_5 S_5})_{ab}^{-1} W_{S_5^b \o}
&= (\Phi M)_{ac} \bigl(M W_{S_5 S_5} M\bigr)_{cd}^{-1} M_{de}W_{S_5^e \o}  \cr
&= -2n_f (M\Phi)_{ac} \Bigl(-(M\Phi) + (2n_f)^2 \chi(0) {\bf 1}_{00}
\Bigr)_{c0}^{-1} \chi(0) \cr
&= 2n_f \chi(0) \Bigl(1 - (2n_f)^2 \chi(0)
(M\Phi)_{00}^{-1}\Bigr)^{-1} \d_{a0} \cr
&= -2n_f \C_{QQ}^{-1}~ \d_{a0} \cr
}}
where in the final step we have used the identification \KK\ .
Similarly,
\eqn\MMMeq{\eqalign{
\Phi_{ac} (W_{S_5 S_5})_{cd}^{-1} \Phi_{db}
&= (\Phi M)_{ac} \bigl(M W_{S_5 S_5} M\bigr)_{cd}^{-1} (M\Phi)_{db} \cr
&= (M\Phi)_{ac} \Bigl(-(M\Phi) + (2n_f)^2 \chi(0) {\bf 1}_{00}
\Bigr)_{cd}^{-1} (M\Phi)_{db} \cr
&= -(M\Phi)_{ab} + (2n_f)^2 \C_{QQ}^{-1} ~\d_{a0} \d_{b0} \cr
}}

This establishes the required results. Substituting \LLLeq\ and \MMMeq\
into \HHHeq\ and \JJJeq\ we find the decay formula
\eqn\NNNeq{
\Phi_{ab} C^{T b\a} ~\hat \C_{\eta^\a A^\l A^\r}
+ 2n_f \C_{QQ}^{-1} ~\hat \C_{G A^\l A^\r}
= a_{\rm em}^a {\a\over\pi}
}
where the decay constants satisfy

\eqn\OOOeq{
f^{a\a} m_{\a\b}^2 f^{T \b b} = 
-(M\Phi)_{ab} + (2n_f)^2 \C_{QQ}^{-1} ~\d_{a0} \d_{b0} 
}

The final step is to identify the 1PI vertices with the couplings
defined in section 1, viz.
\eqn\PPPeq{
\hat \C_{\eta^\a A^\l A^\r} =  g_{\eta^\a \c\c}
}
It is at this point that the central dynamical assumption is made.
In fact, eqs \NNNeq\ and \OOOeq\ are exact identities, following
simply from the definitions and the zero-momentum chiral Ward identities.
To make contact with the radiative decays of the physical particles,
we must assume in particular that the 1PI vertex {\it evaluated
at $k=0$} accurately approximates the physical coupling, which is 
defined {\it on mass-shell}.\foot{The assumption that the 1PI vertices as 
defined here can be identified with the decay couplings of the physical
particles at all rests on the assumption that the dominant particle
poles in the pseudoscalar propagator matrix are indeed those of the $\eta^\a$
(see \FFFeq\ ).} This requires that $\hat\C_{\eta^\a A^\l A^\r}$ has
only a weak momentum dependence in the range $0 \le k^2 \le m_{\eta^\a}^2$.
This is reasonable, since it is defined to be a pole-free, amputated
dynamical quantity. However, as in standard PCAC, the assumption is expected
to be excellent for the $\pi$ but progressively worse as the mass
of the pseudo-Goldstone bosons increases. The hope here, in common
with all attempts to include the $\eta'$ in the framework of PCAC (including
chiral Lagrangians with $1/N_c$ effects included\refs{\KL, \Barc}, is that the
approximation remains sufficiently good at the mass of the $\eta'$.

\vskip0.2cm
It is also important to determine the behaviour of all the quantities 
appearing in these formulae under the renormalisation group. Recall that RG
behaviour was a key factor in the conjecture that $g_{G\c\c}$
may be neglected in first approximation in the decay formula \Aeq\ .
All the required formulae are given in ref.\refs{\Szuoz}. The result is 
that {\it all} the quantities appearing in the final formulae \Aeq\ , \Beq\
(or alternatively \NNNeq\ , \OOOeq\ ) are RG invariant. However, notice
that this is only true for the 1PI vertices evaluated at $k=0$ (and in
fact also on-shell), not for arbitray momenta. The proof is quite
intricate, but since everything can be read off from ref.\refs{\Szuoz}
(see also \refs{\NSVgtmass}) we will not give any further explanation here.

\newsec{Effective Action}

The results in the previous section can be summarised by writing an 
explicit form for the effective action $\C[Q,\phi_5^a]$ compatible
with all the zero-momentum anomalous chiral Ward identities. 
Of course this adds no new physics, but allows the essential results
to be read off in a perhaps simpler and more systematic way.
The resulting effective action is essentially identical to the
di Vecchia--Veneziano\refs{\DiVV}, Rosenzweig--Schechter--Trahern\refs{\RST} 
Lagrangian (with $\o = 0$) though without explicit reference to the $1/N_c$ 
expansion. Less obviously, it is also very closely related to the (non-linear) 
chiral Lagrangians which incorporate the $\eta'$ in the framework of 
large $N_c$\refs{\KL, \Barc}.

The simplest effective action $\C[Q,\phi_5^a]$ compatible with the identities
\IIieq\ , \IIiieq\ has been written down in ref.\refs{\NSVgtmass}. It is
\eqn\aeq{\eqalign{
\C[Q,\phi_5^a] ~=~\int dx~\biggl[&{1\over2A} Q^2 ~+~ BQ Q_{\rm em} ~+~ 
2n_f Q \Phi_{0a}^{-1} \phi_5^a ~+~ a_{\rm em}^a Q_{\rm em} \Phi_{ab}^{-1}
\phi_5^b \cr
&+ {1\over2} \phi_5 \Phi^{-1} f \bigl(-\pl^2 -\m^2\bigr) f \Phi^{-1} \phi_5
\biggr] \cr
}}
The final term is written in matrix notation. $f^{a\a}$ and $\m_{\a\b}^2$
are matrices, where $\m_{\a\b}^2$ is {\it defined} by the Dashen formula
\eqn\beq{
f^{a\a} \m_{\a\b}^2 f^{T\b b} = -M_{ac} \Phi_{cb}
}
The decay constants will subsequently be identified with those in
section 3 (as will the constant $A$). $\m^2$ is of course the pseudo-Goldstone 
boson mass matrix in the OZI limit of QCD, i.e.~before including the
coupling to the gluonic anomaly operator $Q$.

In \aeq\ the simplest choice of kinetic terms for the fields $\phi_5^a$
has been made, with the $f^{a\a}$ chosen to be constants. This is where
the dynamical, pole-dominance, assumption of standard PCAC (or chiral
Lagrangians) is built in. No kinetic terms are included for the composite
operator $Q$ (no glueball poles), and no higher order terms in $Q$ 
are included (these would be suppressed for  large $N_C$).

We have also included terms in \aeq\ involving $Q_{\rm em}(A)$ to
satisfy the anomalous chiral Ward identities with the additional
electromagnetic contribution. The term coupling $Q_{\rm em}$ to 
$\phi_5^a$ is required, whereas the term $Q Q_{\rm em}(A)$ is permitted.
(Like $A$, $B$ is taken to be a constant.) Differentiating \aeq\ , we
immediately obtain
\eqn\ceq{
\Phi_{ab} ~\hat \C_{\phi_5^b A^\l A^\r} = a_{\rm em}^a {\a\over\pi}
}
as previously found in \CCCeq\ .

The second derivatives of $\C[Q,\phi_5^a]$ are
\eqn\deq{
\left(\matrix{\C_{QQ} &\C_{Q\phi_5^b}\cr \C_{\phi_5^a Q} &\C_{\phi_5^a
\phi_5^b} \cr}\right) ~~=~~ \left(\matrix{A^{-1} &2n_f \Phi_{0b}^{-1} \cr
2n_f \Phi_{a0}^{-1} & \Phi^{-1} f \bigl(k^2-\m^2\bigr) f \Phi^{-1}\cr
}\right)
}
which clearly satisfy the identities \IIieq\ and \IIiieq\ . 
We confirm the identification $\C_{QQ}^{-1} = A$. The corresponding
Green functions are found by inversion:
\eqn\eeq{\eqalign{
W_{\o\o} ~~&=~~ -A~ \tilde\D^{-1} \cr
W_{\o S_5^b} ~~&=~~ 2n_f A~ \D_{0d}^{-1}~ \Phi_{db} \cr 
W_{S_5^a \o} ~~&=~~ 2n_f A~ \Phi_{ac}~
\Bigl(f\bigl(k^2-\m^2\bigr)f\Bigr)_{c0}^{-1} \tilde\D^{-1} \cr
W_{S_5^a S_5^b} ~~&=~~ - \Phi_{ac}~ \D_{cd}^{-1}~ \Phi_{db} \cr 
}}
where
\eqn\feq{
\tilde \D ~=~ 1 - (2n_f)^2 A \Bigl(f\bigl(k^2-\m^2\bigr)f\Bigr)_{00}^{-1}
}
and
\eqn\geq{
\D ~=~ f\bigl(k^2-\m^2\bigr)f - (2n_f)^2 A~ {\bf 1}_{00}
}

However, in this form the propagator matrix is clearly not diagonal and
the operators are not normalised so as to couple with unit decay constants
to the physical states. It is therefore convenient to make a change of
variables in $\C$ so that it is written in terms of operators which are
more closely identified with the physical states. Of course this change 
of variables is precisely that described already in section 3.
We therefore define
\eqn\heq{\eqalign{
G &=  Q - W_{\o S_5^a} (W_{S_5 S_5})_{ab}^{-1} \phi_5^b \cr
&= Q + 2n_f A \Phi_{0b}^{-1} \phi_5^b \cr
}}
using \eeq\ for the propagators (c.f.~the identity \LLLeq\ ), and
\eqn\ieq{
\eta^\a = f^{T\a a} \Phi_{ab}^{-1} \phi_5^b
}

In terms of these operators, the effective action is
\eqn\jeq{\eqalign{
\C[G,\eta^\a] ~=~ \int dx~ \biggl[ &{1\over 2A} G^2 ~+~ BGQ_{\rm em} ~+~
a_{\rm em}^a f_{a\a}^{-1} \eta^\a Q_{\rm em} ~-~ 
2n_f AB f_{0\a}^{-1} \eta^\a Q_{\rm em} \cr
&+ {1\over2} \eta\Bigl[(-\pl^2 -\m^2) - (2n_f)^2 A f^{T -1} {\bf 1}_{00} f^{-1}
\Bigr] \eta ~\biggr] \cr
}}
It is then straightforward to read off the propagators
\eqn\keq{\eqalign{
\la G~G\ra &= -  A \cr
\la \eta^\a ~\eta^\b\ra &= {-1\over k^2 - m_{\eta^\a}^2}\d^{\a\b} \cr
}}
where the diagonal mass matrix for the physical $\eta^\a$ states 
satisfies the generalised Dashen fomula
\eqn\leq{
f^{a\a} m_{\a\b}^2 f^{T\b b} = f^{a\a} \m_{\a\b}^2 f^{T\b b}
+ (2n_f)^2 A \d_{a0} \d_{b0}
}
Of course this is identical to \Beq\ . This confirms the identification of the 
decay constants in the effective action with those in sections 1 and 3.

Finally, to obtain the decay formula itself, we take functional
derivatives of $\C$ to get
\eqn\meq{\eqalign{
\hat \C_{G A^\l A^\r} &= B \cr
f^{a\a} \hat \C_{\eta^\a A^\l A^\r} &=  a_{\rm em}^a {\a\over\pi} 
- 2n_f AB \d_{a0} \cr
}}
Combining these, we find
\eqn\neq{
f^{a\a} ~\hat \C_{\eta^\a A^\l A^\r} ~+~ 2n_f A ~\hat \C_{G A^\l A^\r} ~=~
a_{\rm em}^a {\a\over\pi}
}
in agreement with \Aeq\ . Notice the importance of including the $QQ_{\rm em}$
coupling in \aeq\ in obtaining this result.

\newsec{$U_A(1)$ PCAC}

For the third of our variations on a theme, we present a derivation of
the decay formula following as closely as possible the traditional 
language of PCAC. This should be reasonably self-contained, though 
we will use the compact notation defined at the start of section 2 
and simply quote the chiral Ward identities without proof.

Consider first QCD by itself without the coupling to electromagnetism.
The axial anomaly equation is
\eqn\aaeq{
\pl^\m J_{\m5}^a = M_{ab}\phi_5^a + 2n_f Q \d_{a0}
}
where $J_{\m5}^a$ is the axial current, $\phi_5^a$ the pseudoscalar
quark bilinear operator and $Q$ the topological charge. $M_{ab}$
describes the quark masses and $\Phi_{ab}$ the condensates.
The anomalous chiral Ward identities, at zero momentum, for the 
propagators (i.e.~two-point Green functions) of these operators are
\eqn\bbeq{\eqalign{
&2n_f \la Q~Q\ra \d_{a0} + M_{ac} \la \phi_5^c~Q\ra = 0 \cr
&2n_f \la Q~\phi_5^b\ra \d_{a0} + M_{ac}\la\phi_5^c~\phi_5^b\ra 
+ \Phi_{ab} = 0 \cr
}}
which imply
\eqn\cceq{
M_{ac} M_{bd} \la \phi_5^c~\phi_5^d\ra = - (M\Phi)_{ab} 
+ (2n_f)^2 \la Q~Q\ra \d_{a0}\d_{b0}
}
We also need the result for the general form of the topological 
susceptibility:
\eqn\ddeq{
\chi(0) \equiv \la Q~Q \ra = {-A\over 1 - (2n_f)^2 A (M\Phi)_{00}^{-1} }
}

Although the pseudoscalar operators $\phi_5^a$ and $Q$ indeed couple to the 
physical states $\eta^\a = \eta', \eta, \pi^0$, it is more convenient to
redefine linear combinations such that the resulting propagator matrix
is diagonal and properly normalised. That is, we define operators
$\eta^\a$ and $G$ such that
\eqn\eeeq{
\left(\matrix{\la Q~Q \ra &\la Q~\phi_5^b \ra \cr
\la \phi_5^a ~Q \ra &\la \phi_5^a~\phi_5^b \ra }\right) ~~\rta~~
\left(\matrix{\la G~G\ra &0 \cr
0 &\la \eta^\a ~ \eta^\b \ra }\right)
}
This is achieved by 
\eqn\ffeq{\eqalign{
G &=  Q - \la Q~\phi_5^a\ra (\la \phi_5 ~\phi_5\ra)_{ab}^{-1} \phi_5^b \cr
&= Q + 2n_f A \Phi_{0b}^{-1} \phi_5^b \cr
}}
and
\eqn\ffFeq{
\eta^\a = f^{T\a a} \Phi_{ab}^{-1} \phi_5^b 
}
With this choice, the $\la G~G\ra$ propagator is
\eqn\ggeq{
\la G~G\ra = - A
}
and we impose the normalisation
\eqn\hheq{
\la \eta^\a ~\eta^\b\ra = {-1\over k^2 - m_{\eta^\a}^2}\d^{\a\b}
}
This implies that the constants $f^{a\a}$ in \ffFeq\ , which we see shortly
are simply the decay constants, must satisfy the (Dashen) identity
\eqn\iieq{\eqalign{
f^{a\a} m_{\a\b}^2 f^{T\b b} &= \Phi_{ac} (\la \phi_5~\phi_5 \ra)_{cd}^{-1}
\Phi_{db} \cr
&= -(M\Phi)_{ab} + (2n_f)^2 A \d_{a0}\d_{b0} \cr
}}
The last line follows from the Ward identities \cceq\ and \ddeq\ .
In terms of these new operators, the anomaly equation \aaeq\ now
reads simply:
\eqn\jjeq{
\pl^\m J_{\m5}^a = f^{a\a} m_{\a\b}^2 \eta^\b + 2n_f G \d_{a0}
}

After these preliminaries, we now recall how conventional PCAC is
applied to the calculation of $\pi^0\rta\c\c$. The pion decay constant
is defined as the coupling of the pion to the axial current
\eqn\kkeq{
\la 0|J_{\m5}^3|\pi\ra = ik_\m f_\pi ~~~~\Rightarrow~~~~
\la 0|\pl^\m J_{\m5}^3|\pi\ra = f_\pi m_\pi^2
}
and satisfies the Dashen formula
\eqn\lleq{
f_\pi^2 m_\pi^2 = -(m_u + m_d)\la \bar q q\ra
}
The next step is to define a `phenomenological pion field' $\pi$ by
\eqn\mmeq{
\pl^\m J_{\m5}^3 \rta f_\pi m_\pi^2 \pi
}
This is the step at which the crucial `pole-dominance' assumption is made.
Now include electromagnetism. The full anomaly equation is extended as in
\Deq\ to include the $F^{\m\n} \tilde F_{\m\n}$ contribution. Using \mmeq\
we therefore have
\eqn\nneq{\eqalign{
ik^\m\la \c\c|J_{\m5}^3|0 \ra &= f_\pi m_\pi^2 \la \c\c|\pi|0\ra
+ a_{\rm em}^a {\a\over8\pi} \la \c\c|F^{\m\n} \tilde F_{\m\n}|0\ra \cr
&= f_\pi m_\pi^2 \la \pi~\pi\ra \la \c\c|\pi\ra
+ a_{\rm em}^a {\a\over8\pi} \la \c\c|F^{\m\n} \tilde F_{\m\n}|0\ra \cr
}}
where $\la \pi~\pi\ra$ is the pion propagator $-1/(k^2-m_\pi^2)$.
At zero momentum, the l.h.s.~vanishes because of the explicit $k_\m$
factor and the absence of massless poles. We therefore find, defining the
couplings as in \Ceq\ ,
\eqn\ooeq{
f_\pi g_{\pi\c\c} = a_{\rm em}^3 {\a\over\pi}
}

In the full theory including the flavour singlet sector and the gluonic
anomaly, we find a similar result. The `phenomenological fields'
are defined by \jjeq\ where the decay constants satisfy the generalised
Dashen formula \iieq\ . Notice, however, that they are {\it not} simply
related to the couplings to the axial current as in \kkeq\ for the
flavour non-singlet. We therefore find:
\eqn\ppeq{\eqalign{
ik^\m\la \c\c|J_{\m5}^a|0 \ra &= f^{a\a} m_{\a\b}^2 \la \c\c|\eta^\b|0\ra
+ 2n_f \la \c\c|G|0\ra \d_{a0}
+ a_{\rm em}^a {\a\over8\pi} \la \c\c|F^{\m\n} \tilde F_{\m\n}|0\ra \cr
&= f^{a\a} m_{\a\b}^2 \la \eta^\b~\eta^\c \ra \la \c\c|\eta^\c \ra
+ 2n_f \la G~G\ra \la \c\c|G\ra \d_{a0}
+ a_{\rm em}^a {\a\over8\pi} \la \c\c|F^{\m\n} \tilde F_{\m\n}|0\ra \cr
}}
using the fact that the propagators are diagonal in the basis $\eta^\a, G$.
Using the explicit expressions \ggeq\ and \hheq\ for the propagators,
evaluating at zero momentum, and setting the l.h.s.~to zero, we find
in this case:
\eqn\qqeq{
f^{a\a} ~g_{\eta^a\c\c} + 2n_f A ~g_{G\c\c} ~\d_{a0} ~=~ a_{\rm em}^a 
{\a\over\pi}
}
where the extra coupling $g_{G\c\c}$ is defined through \ppeq\ .
This completes the `$U_A(1)$ PCAC' derivation. It is evidently a
straightforward generalisation of conventional PCAC with the 
necessary modification of the usual formulae to take account of the
extra gluonic contribution to the axial anomaly in the flavour
singlet channel, the key point being the identification of the 
operators $\eta^\a$ and $G$ in \jjeq\ .

\newsec{Acknowledgements}

I would like to thank S.~Narison and G.~Veneziano for helpful discussions.
This work was supported in part by PPARC grant GR/L56374, and by
the EC TMR Network grant FMRX-CT96-0008.

\listrefs
\bye